\documentclass[11pt]{article}
\usepackage{epsfig}
\usepackage{amsmath}

\newlength{\dinwidth}
\newlength{\dinmargin}
\newcommand{\tdm}[1]{\mbox{\boldmath $#1$}}
\newcommand{\as}{\alpha_s}
\def\gV{\gamma V}
\def\imag{\mathcal{I}\mathrm{m}\,}

\begin{document}

\titlepage

\begin{center}
\vspace*{2cm}
{\Large \bf
Diffractive heavy vector meson production\\
from the BFKL equation
}

\vspace*{1cm}
{\large Rikard Enberg$^{a}$, Leszek Motyka$^{a,c}$ and
Gavin Poludniowski$^{b}$} \\
\vspace*{1cm}
{\small
$^{a}$ Department of Radiation Sciences, Uppsala University, Box 535,
       S-751 21 Uppsala, Sweden   \\
$^{b}$ Department of Physics and Astronomy, University of Manchester,
       Manchester M13 9PL, UK \\
$^{c}$ Institute of Physics, Jagellonian University, Reymonta 4,
       30-059 Krak\'{o}w, Poland  \\
}
\end{center}

\vspace*{0.5cm}

\begin{abstract}
Diffractive heavy vector meson photoproduction accompanied by proton
dissociation is studied for arbitrary momentum transfer.
The process is described by the non-forward BFKL equation, for which
a complete analytical solution is found, giving the scattering
amplitude. The impact of non-leading corrections to the BFKL equation
is also analysed. Results are compared to the HERA data on
$J/\psi$ production.
\end{abstract}

%%%%%%%%%%%%%%%%%%%%%%%%%%%%%%%%%%%%%%%%%%%%%%%%%%%%%%%%%%%%%%%%%%%%%%%%%%%
\section{Introduction}\label{sec1}

Quantum Chromodynamics offers unique opportunities to study the
richness of dynamical phenomena of nonlinear quantum field theory.
One of the most interesting problems is related to
the colour flow in high energy scattering. In particular, diffractive
processes correspond to an exchange of a colour singlet system of
quarks and gluons between scattering objects. Such diffractive phenomena
possess a very clean experimental signature, namely a large rapidity
interval devoid of particles (i.e.\ a rapidity gap).

The perturbative QCD description of the hard colour singlet exchange across
a large rapidity interval $y$ relies on the Balitsky-Fadin-Kuraev-Lipatov
(BFKL) equation \cite{BFKL,Lipatov}. In this framework, the leading powers of
rapidity in the perturbative expansion are resummed, giving the amplitude
for hard pomeron exchange. The pomeron is viewed as a composite system of
two reggeized gluons in the colour singlet state. The status of the BFKL
approach to QCD amplitudes is still under discussion, and both theoretical
improvements and experimental tests are necessary.

\begin{figure}
\begin{center}
\epsfig{width=0.45 \columnwidth, file=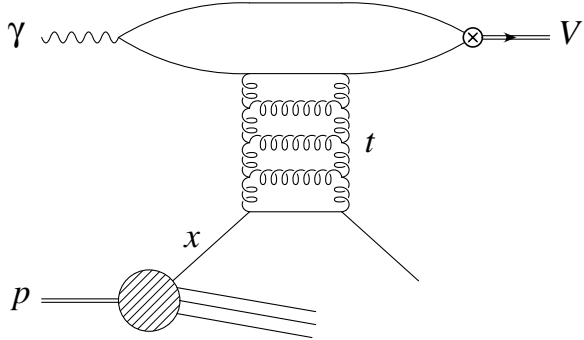} \\
\end{center}
\caption{\em Feynman diagram illustrating vector meson
photoproduction at high momentum transfer.}
\label{fig1}
\end{figure}

Diffractive photoproduction of a heavy vector meson, separated from
the proton remnant by a large rapidity gap has been proposed \cite{FR,BFLW}
as an ideal probe of the BFKL pomeron, see fig.\ \ref{fig1}.
Indeed, this process permits detailed studies of both the momentum
transfer and the rapidity dependence of the scattering amplitude.
The vector meson mass and the momentum transfer
$t$ provide the hard scale required for the perturbative treatment of
QCD processes, and the sensitivity to the infra-red region
is small, contrary to the case of inclusive hard diffraction.

There are some very recent measurements of this process from HERA
\cite{ZEUS}
that allow the theoretical models to be tested.
The available calculations \cite{FR,BFLW,FP} of the cross-sections for heavy
vector meson production are based on the Mueller-Tang approximation \cite{MT}
to the solution of the leading logarithmic BFKL equation.
In this approximation, parts of the amplitude which
vanish in the limit $y \to \infty$ are neglected \cite{MMR,BFLW}.

Recently, it has been shown \cite{FP} that fitting these results to the data,
one obtains a good quantitative agreement with the differential cross-section.
Still, the important non-leading corrections to the BFKL kernel
\cite{BFKLNL} are not accounted for.
Furthermore, the Mueller-Tang approximation is only good for
very large rapidities, and may need improvement in order to
understand the experimental data. Indeed, it has been found that to
describe the events with gaps between jets
subleading corrections to the  Mueller-Tang picture are
important \cite{MMR,EIM}.

Thus, the main goal of this paper is to investigate diffractive
heavy\footnote{Light vector meson production
will be studied in a forthcoming paper \cite{LVM}.}
vector meson photoproduction beyond the leading logarithmic BFKL equation and
beyond the Mueller-Tang approximation. The obtained results are
compared to previous ones \cite{FR,BFLW} and to the
experimental data from HERA \cite{ZEUS}.
In Sec.\ \ref{sec2} we define the framework, in Sec.\ \ref{sec3} the 
BFKL equation is
presented, and in Sec.\ \ref{sec4} an exact solution of the equation
is derived. Properties of the exact and numerical
solutions are studied Sec.\ \ref{sec5}, comparison with data is performed
in Sec.\ \ref{sec6}, and in Sec.\ \ref{sec7} conclusions are given.

%%%%%%%%%%%%%%%%%%%%%%%%%%%%%%%%%%%%%%%%%%%%%%%%%%%%%%%%%%%%%%%%%%%%%%%%%%%
\section{Hard colour singlet exchange}\label{sec2}

The diffractive process $\gamma p \to V X$ at large momentum
transfer $t$ (see fig.\ \ref{fig1}) takes place by exchange of the BFKL pomeron.
It has been demonstrated, that at large momentum transfer, the hard pomeron
couples predominantly to individual partons in the proton \cite{BFLLR}.
Thus, the cross-section may be factorized into a product of the parton
level cross-section and the parton distribution functions,
\begin{align}
\frac{d\sigma (\gamma p \rightarrow VX)}{dt\, dx_j} \;=\;
\biggl(
\frac{4N_{c}^{4}}{(N_{c}^{2}-1)^2}
%\left( {C_A \over C_F} \right) ^2
G(x_j,t)+
\sum_{f}[q_{f}(x_j,t)+\bar{q}_{f}(x_j,t)]\biggr)\;
\frac{d\sigma (\gamma q \rightarrow Vq)}{dt},
\label{dsdtgp}
\end{align}
where $N_{c}=3$, $G(x_j,t)$ and $q_{f}(x_j,t)$ are the gluon and quark
distribution functions respectively, and  $W^2$ is the $\gamma p$
centre-of-mass energy squared.
The struck parton in the proton, initiating a jet in the proton hemisphere,
carries the fraction $x_j$ of the longitudinal momentum of the incoming
proton. The partonic cross-section, characterized by the invariant
collision energy squared $\hat s = x_j W^2$ is expressed in terms of
the amplitude  $\mathcal{A}(\hat{s},t)$,
\begin{equation}
\frac{d \sigma}{dt} = \frac{1}{16 \pi} |\mathcal{A}(\hat s,t)|^2.
\label{dsdt}
\end{equation}
The amplitude is dominated by its imaginary part,
which we shall parametrize, as in \cite{FR,FP},
by a dimensionless quantity  $\mathcal{F}$
\begin{equation}
\imag \mathcal{A}(\hat s,t) =
{16 \pi
%\alpha_s^2
\over 9 t^2} \mathcal{F}(z,\tau)
\label{FA}
\end{equation}
where $z$ and $\tau$ are defined by
\begin{align}
z &\;=\; \frac{3\alpha_{s}}{2\pi}
\ln \biggl( \frac{\hat s}{\Lambda^{2}} \biggr)
\label{zdef}
\\
\tau &\;=\; \frac{|t|}{M_{V}^{2}+ Q_{\gamma}^{2}},
\label{taudef}
\end{align}
where
$M_{V}$ is the mass of the vector meson, $Q_\gamma$ is the photon
virtuality\footnote{In this paper we only consider $Q_\gamma=0$.}
and $\Lambda^{2}$ is a characteristic mass scale related to
$M_V^2$  and $|t|$.
Following the results of \cite{FP} we assume
$\Lambda^2 = M_V ^2 + Q_\gamma^2 $.
For completeness, we give the cross-section expressed
in terms of $\mathcal{F}(z,\tau)$, where the real part of the
amplitude is neglected,
\begin{equation}
\frac{d\sigma (\gamma q \rightarrow Vq)}{dt} \; = \;
\frac{16\pi}{81 t^4}
%\frac{\alpha_{s}^{4}}{t^{4}}\;
|\mathcal{F}(z,\tau)|^{2}.
\label{dsdtgq}
\end{equation}
This representation is rather convenient for the calculations
performed in Sec.\ \ref{sec4}.

%%%%%%%%%%%%%%%%%%%%%%%%%%%%%%%%%%%%%%%%%%%%%%%%%%%%%%%%%%%%%%%%%%%%%%%%%%%
\section{The BFKL equation}\label{sec3}

The imaginary part $\imag A(\hat s,t)$ of the  amplitude for the process
$\gamma p \rightarrow V + {\mathrm{gap}}+ X + {\mathrm{jet}}$
corresponds to the diagram in fig.\ \ref{fig1} illustrating QCD pomeron
exchange, and can be written in the following form:
\begin{equation}
\imag A(\hat s,t=-q^2) = \int {d^2\tdm k\over \pi}
{\Phi^{0}_{\gV} (k^2, q^2) \Phi_{qq} (x,\tdm k,\tdm q)
\over
[(\tdm k + \tdm q /2)^2 +s_0][(\tdm k - \tdm q /2)^2+s_0]}.
\label{ima}
\end{equation}
In this equation, $x$ is the longitudinal momentum fraction of the
incoming proton taken by the hard pomeron,  $x = \Lambda^2 / \hat s$,
$\tdm q/ 2 \pm \tdm k$ denote the transverse  momenta of the exchanged
gluons, and $\tdm q$ is the transverse part of the
momentum transfer.  In the propagators corresponding to the
exchanged gluons we include the parameter $s_0$ which can be
viewed as the effective representation of the inverse of
the colour confinement radius squared.
%%% comm 3a%%%
Investigation of such non-perturbative effects in the
gluon propagator at low virtualities was performed,
for instance, in \cite{GLUs0}, where it was found
that $0.1$~GeV$^2 < s_0 < 0.5$~GeV$^2$.
Recent lattice studies \cite{Latt1} indicate
a value of $s_0$ between $0.25$~GeV$^2$ and $0.65$~GeV$^2$. 
Having some freedom here, we choose $s_0 = 0.5$~GeV$^2$.
%%%
The sensitivity of the cross-section to the magnitude of $s_0$
can serve as an estimate of the sensitivity of the results to the
contribution coming from the infra-red region. It should be noted that
formula (\ref{ima}) gives a finite result in the limit $s_0=0$~GeV$^2$. 
%%% comm 3c %%%
An interesting discussion of non-perturbative effects in the gluon propagator 
in the context of vector meson production can be found in \cite{DucatiSauter}.
%%%

The couplings of the external particle pair to the colour singlet
gluonic ladder are described, in the high energy limit, by impact factors
$\Phi_{\gV} ^0 (k^2, q^2)$ and $\Phi_{qq} ^0 (k^2, q^2)$
for the $\gamma \to V$ transition and the quark elastic
scattering, respectively.
The impact factors are obtained in the perturbative QCD framework
and we approximate them by the leading terms in the perturbative
expansion \cite{Ryskin}:
\begin{align}
\Phi^0 _{\gV}\;=\;
\frac{{\mathcal C} \as}{2}\, \biggl(\frac{1}{\bar{q}^{2}}-
\frac{1}{q_{\|}^{2}+k^{2}} \biggr),
\nonumber \\
\Phi^0 _{qq}\;=\; \as.
\hspace{2.5cm}
\label{impfmom}
\end{align}
In the former formula, factorization of the
scattering process and the meson formation is assumed, and the
non-relativistic approximation of the meson wave function is used.
In this approximation the quarks in the meson have collinear four-momenta and
$M_{V}=2M_{q}$ where $M_{q}$ is the mass of the constituent quark.
To leading order accuracy, the constant $\mathcal C$ may be related
to the vector meson leptonic decay width
\begin{align}
\mathcal{C}^{2}\;=\;\frac{3\Gamma_{ee}^{V}M_{V}^{3}}{\alpha}.
\end{align}
We have also defined
\begin{align}
\bar{q}^{2}\;=\;q_{\|}^{2}+q^{2}/4, \\
q_{\|}^{2}\;=\;(Q_{\gamma}^{2}+M_{V}^{2})/4.
\label{C}
\end{align}
%
% comm 1
This standard approximation is based on the fact, that
the typical quark three-momenta in charmonium states are
much smaller than the charm quark mass. Still, these
momenta are non-zero, leading to a smearing of the
non-relativistic form-factor. The resulting  corrections were
extensively discussed e.g.\ in \cite{FKS} for the forward
scattering case.
Moreover, it is expected at larger momentum transfer
$|t| > M_V^2$, that the relativistic effects become enhanced,
as the mass is no longer the largest scale in the process.
For instance, in the non-relativistic approximation
one neglects helicity flip processes, whose relative
contribution may grow with increasing momentum transfer,
as was demonstrated for a single helicity flip amplitude
for light vector mesons \cite{IKSS}.
Therefore, the applied approximation may be
inaccurate at very large momentum transfers.

The function $\Phi_{qq}(x,\tdm k,\tdm q)$ satisfies the BFKL equation, which
in the leading $\ln(1/x)$ approximation has the following form:
$$
\Phi_{qq}(x,\tdm k,\tdm q)=\Phi^0 _{qq}(k^2, q^2)+
{3\alpha_s \over 2\pi^2}
\int_x^1{dx^{\prime}\over x^{\prime}} \int
{d^2\tdm k' \over (\tdm k' - \tdm k)^2 + s_0} \times
$$
$$
\left\{\left[{{\tdm k_1^2}\over {\tdm k_1^{\prime 2}} + s_0}   +
{{\tdm k_2^2}\over {\tdm k_2^{\prime 2}} + s_0} - q^2
 {(\tdm k' - \tdm k)^2+s_0 \over ({\tdm k_1^{\prime 2}} + s_0)
 ({\tdm k_2^{\prime 2}} + s_0)}
\right] \Phi_{qq}(x',\tdm k' ,\tdm q) - \right.
$$
\begin{equation}
\left. \left[{{\tdm k_1^2}\over {\tdm k_1^{\prime 2}}  +
(\tdm k' - \tdm k)^2 +2s_0} +
{{\tdm k_2^2}\over {\tdm k_2^{\prime 2}}  +
(\tdm k' - \tdm k)^2 +2s_0} \right]
\Phi_{qq}(x',\tdm k,\tdm q) \right\}
\label{bfkl}
\end{equation}
where
\begin{equation}
{\tdm k_{1,2}} = {\tdm q \over 2}\pm \tdm k, \qquad
{\tdm k_{1,2}^{\prime}} = {\tdm q \over 2} \pm \tdm k^{\prime}
\label{k12}
\end{equation}
denote the transverse momenta of the gluons. At leading logarithmic accuracy,
a fixed value of the QCD coupling
$\alpha_s$ should be used in equations (\ref{impfmom})
and (\ref{bfkl}).

It is known that the BFKL equation can acquire significant
non-leading contributions \cite{BFKLNL}.
Although the structure of those corrections is fairly complicated,
their dominant part is rather simple, and follows from restricting
the integration region in the real emission term in
equation (\ref{bfkl}) \cite{KC,COLLINEAR}.
For $q=0$ the relevant limitation is
\cite{KC,KMSTAS}
\begin{equation}
{k'}^2 \le k^2 {x'\over x}.
\label{kc1}
\end{equation}
This follows from  the requirement that the virtuality of the
gluons exchanged along the chain is dominated by the transverse
momentum squared. The  constraint (\ref{kc1}) can be shown to
exhaust about $70 \%$ of the next-to-leading corrections
to the QCD pomeron intercept \cite{BFKLNL,KC}.
Generalization of the constraint (\ref{kc1}) to the case of
a non-forward configuration with $q^2 \ge 0$ is assumed to
take the following form
\cite{PSIPSI,EIM}:
\begin{equation}
{k'}^2 \le (k^2+ q^2/4) {x'\over x}.
\label{kc2}
\end{equation}
%%% comm 2
The latter formula gives at $k^2 < q^2 /4$ a less
restrictive relation for $k'$ in respect to $k$, than
constraint  (\ref{kc1}) in the forward case.
Consequently, the generalized constraint
(\ref{kc2}) may realize, at larger momentum transfer,
a smaller fraction of the non-leading corrections
than the quoted 70\% at $q^2 = 0$. This would mean
that our estimate of the cross-section could be too high
at larger $q^2$.
%%%

Another important part of the non-leading corrections to
the BFKL equation is related to running of the coupling
constant within the ladder. To be consistent, the
running coupling will also be used in the impact factors
(\ref{impfmom}). Besides the BFKL equation (\ref{bfkl})
in the leading logarithmic approximation we shall therefore
%also consider the equation which will embody the constraint
also consider the equation embodying the constraint
(\ref{kc2}) and a running coupling  in order to estimate
the effects of the non-leading contributions.

The corresponding equation which contains constraint (\ref{kc2}) in the
real emission term reads:
$$
\Phi_{qq}(x,\tdm k,\tdm q)=\Phi_{qq} ^0(k^2, q^2)+
{3\alpha_s(\mu^2)\over 2\pi^2}
\int_x^1{dx^{\prime}\over x^{\prime}} \int
{d^2\tdm k' \over (\tdm k' - \tdm k)^2 + s_0} \times
$$
$$
\left\{\left[{{\tdm k_1^2}\over {\tdm k_1^{\prime 2}} + s_0}   +
{{\tdm k_2^2}\over {\tdm k_2^{\prime 2}} + s_0} - q^2
{(\tdm k' - \tdm k)^2+s_0 \over ({\tdm k_1^{\prime 2}} + s_0)
({\tdm k_2^{\prime 2}} + s_0)}
\right]
%\times \right.
%$$
%
%$$
\Phi_{qq}(x',\tdm k' ,\tdm q)
\Theta \left( (k^2+q^2/4)x'/x-k^{\prime 2}) \right)  -
\right.
$$
\begin{equation}
\left. \left[{{\tdm k_1^2}\over {\tdm k_1^{\prime 2}}  +
(\tdm k' - \tdm k)^2 +2s_0} +
{{\tdm k_2^2}\over {\tdm k_2^{\prime 2}}  +
(\tdm k' - \tdm k)^2 +2s_0} \right]
\Phi_{qq}(x',\tdm k,\tdm q) \right\}.
\label{bfklkc}
\end{equation}
with the scale of the coupling set to $\mu^2=k^2+q^2/4+s_0$.
The scales of the coupling constants in the
impact factors should be related to the virtualities
entering the vertices. A natural choice is then
\[
\mu_1 ^2  =  k^2 + M_c^2\qquad \mbox{in} \qquad \Phi^0 _{\gamma V},
\]
\begin{equation}
\mu_2 ^2  =  k ^2 + s_0 \qquad \mbox{in} \qquad \Phi^0 _{qq},
\label{scales}
\end{equation}
with $k^2$ being the virtuality of the gluon
entering the vertex. We will also consider another
choice
\begin{equation}
{\mu_1'} ^2 = \mu_1^2 /4\;\;\; \mbox{and}\;\;\; {\mu_2'} ^2 = \mu_2^2 /4,
\label{scales2}
\end{equation}
necessary in order to obtain a good fit to the data.
A similar choice of scales was needed to describe double-tagged
events at LEP in an analogous NL-BFKL framework \cite{KMDT}.
Equations (\ref{bfkl})
and (\ref{bfklkc}) are solved using an approximate numerical
technique, described in detail in \cite{PSIPSI} and \cite{EIM}.

%%%%%%%%%%%%%%%%%%%%%%%%%%%%%%%%%%%%%%%%%%%%%%%%%%%%%%%%%%%%%%%%%%%%%%%%%%%
\section{The exact LL BFKL solution}\label{sec4}

The BFKL kernel (\ref{bfkl}) in the leading logarithmic approximation
exhibits, in the impact parameter representation, invariance
under conformal transformations \cite{Lipatov}.
The conformal symmetry of the kernel permits the following expansion
of the amplitude in the basis of eigenfunctions $E_{n,\nu}$ \cite{Lipatov}:
\begin{align}
\mathcal{F}(z,\tau)\;=\;
\frac{t^{2}}{(2\pi)^{3}}\sum_{n=-\infty}^{n=\infty}
 \int_{-\infty}^{\infty}
d\nu\; \frac{\nu^{2}+n^{2}/4}{[\nu^{2}+(n-1)^{2}/4][\nu^{2}+(n+1)^{2}/4]}
\nonumber
\\ \times
\exp [\chi_{n}(\nu)z]\; I^{A}_{n,\nu}(q)\,  ({I^{B}_{n,\nu} (q)})^*
\hspace{2cm}
\label{bfklampl}
\end{align}
where
\begin{equation}
\chi_{n}(\nu)\;=\; 4\mathcal{R}e\biggl(\psi(1)-\psi(1/2+|n|/2+i\nu)\biggr)
\end{equation}
is proportional to the eigenvalues of the BFKL kernel and
\begin{equation}
I^{A}_{n,\nu}(q)\; = \; \int \frac{d^{2}k}{(2\pi)^{2}}
\mathcal{I}_{A}(k,q)\int
d^{2}\rho_{1}\,d^{2}\rho_{2}\; E_{n,\nu}(\rho_{1},\rho_{2})
\exp (ik\cdot \rho_{1} +i(q-k) \cdot \rho_{2})
\label{impf}
\end{equation}
(and analogously for the index $B$). The eigenfunctions are given by
\begin{equation}
E_{n,\nu}(\rho_{1},\rho_{2})\;=\;
\biggl(\frac{\rho_{1}-\rho_{2}}{\rho_{1}\rho_{2}}\biggr)^{h}
\biggl(\biggl(\frac{\rho_{1}-\rho_{2}}{\rho_{1}\rho_{2}}\biggr)^{*}\biggr)^{\tilde{h}}
\label{E}
\end{equation}
where $h=1/2+n/2+i\nu$ and $\tilde{h}=1/2-n/2+i\nu$.
Here $k$ and $q$ are transverse two dimensional momentum vectors, and
$\rho_{1}$ and $\rho_{2}$ are position space vectors in the
standard complex representation (e.g.\ $k=k_x + i k_y$).
The scalar product in this representation is
given by, e.g., $k\cdot \rho_{1}=k^{*}\rho_{1}/2+k\rho_{1}^{*}/2$.
The functions $\mathcal{I}_{A} = \Phi^0 _{\gV}$ and
$\mathcal{I}_{B} = \Phi^0 _{qq}$ are the impact factors (\ref{impfmom}).

The quark impact factor in representation (\ref{impf})
was found in \cite{MMR}, generalizing
the Mueller-Tang subtraction \cite{MT} to non-zero conformal spin;
\begin{equation}
I^{qq}_{n,\nu}(q) \;=\;
-\frac{4\pi\,\as\, i^n}{|q|}\;
\biggl(\frac{|q|^{2}}{4}\biggr)^{i\nu}\,
\biggl(\frac{q^*}{q} \biggr) ^{n/2}\,
\frac{\Gamma(1/2+n/2-i\nu)} {\Gamma(1/2+n/2+i\nu)}
\label{iq}
\end{equation}
for even~$n$ and $I^{qq}_{n,\nu}=0$ for odd~$n$.

The impact factor for the $\gamma \rightarrow V$ transition
is known for $n=0$ \cite{BFLW}. We shall generalize this result to
arbitrary $n$, which requires evaluating the
following integrals
\begin{align}
I^{\gV}_{n,\nu}(q)\;=\;\int \frac{d^{2}k}{(2\pi)^{2}} \,
\mathcal{I}_{\gamma V}(k,q)
\int d^{2}\rho_{1}\, d^{2}\rho_{2}\;
E_{n,\nu}(\rho_{1},\rho_{2}) \nonumber \\
\times \exp (ik^{*}\rho_{1}/2 +ik\rho_{1}^{*}/2 +
i(q^{*}-k^{*})\rho_{2}/2 +i(q-k)\rho_{2}^{*}/2).
\end{align}
Changing variables to $\rho_{1}=R+\rho/2$, $\rho_{2}=R-\rho/2$ and
integrating over $d^{2}k$ we get
\begin{align}
I^{\gV}_{n,\nu}(q)\;=\;
-\frac{\mathcal{C}\, \as }{4\pi} \int d^{2}\rho \,
K_{0}(q_{\|}|\rho|)\,
\int d^{2}R\;
E_{n,\nu}(R+\rho/2,R-\rho/2)\,
\exp (iq^{*}R/2 +iqR^{*}/2).
\label{igv}
\end{align}
The integral over $d^2R$ was obtained by Navelet and Peschanski \cite{NP}.
Inserting their result one obtains
\begin{align}
I^{\gV}_{n,\nu}(q)\;=\;
-\frac{\mathcal{C} \,\as}{4\pi}
\int d^{2}\rho\, \frac{(-1)^{n}|\rho|}{2\pi^{2}}\,
b_{n,\nu} \,
\hat{E}_{n,\mu}(\rho,\rho^{*})\,
K_{0}(q_{\|}|\rho|)
\label{igv1}
\end{align}
where
\begin{align}
\hat{E}_{n,\mu}(\rho,\rho^{*})\; = \;
\biggl(\frac{|q|}{8}\biggr)^{2i\nu} \biggl(\frac{q^{*}}{q} \biggr)^{n/2}
\Gamma(1-i\nu+n/2)\, \Gamma(1-i\nu-n/2) \nonumber \\ \times
\biggl[J_{n/2-i\nu}(q^{*}\rho/4)\, J_{-n/2-i\nu}(q\rho^{*}/4)
\,-\,
(-1)^{n} \, J_{-n/2+i\nu}(q^{*}\rho/4)\, J_{n/2+i\nu}(q\rho^{*}/4)  \biggr]
\label{ehat}
\end{align}
and
\begin{align}
b_{n,\nu}\,=\,\frac{2^{4i\nu}\pi^{3}}{|n|/2-i\nu} \;
\frac{\Gamma(|n|/2-i\nu+1/2)\Gamma(|n|/2+i\nu)}
{\Gamma(|n|/2+i\nu+1/2)\Gamma(|n|/2-i\nu)}.
\label{bnu}
\end{align}

Thus, we have a result in terms of a double integral over
$d^{2}\rho=|\rho|\, d|\rho| \, d\phi$.
Further, we represent the Bessel functions by their power series
expansions
$ J_{\sigma}(z) = (z/2)^{\sigma} \sum_{k=0} ^{\infty} (-1)^k \, (z/2)^{2k} /
[\Gamma(k+1) \Gamma(\sigma+k+1)]$
and obtain
\begin{align}
%\biggl[
J_{n/2-i\nu}(q^{*}\rho/4)\, J_{-n/2-i\nu}(q\rho^{*}/4)
\,-\,(-1)^{n}\,J_{-n/2+i\nu}(q^{*}\rho/4)\, J_{n/2+i\nu}(q\rho^{*}/4)
%\biggr]
\;=\; \nonumber \\
\biggl[
%\biggl( \frac{|q||\rho|}{8}\biggr)^{-2i\nu}
\sum_{k=0}^{\infty} \sum_{l=0}^{\infty} \frac{(-1)^{k+l}\,
(|q||\rho|/8)^{2k+2l-2i\nu}\, \exp [i\phi(2l-2k+n)]}
{\Gamma(k+1)\,\Gamma(l+1)\,\Gamma(1+l+n/2-i\nu)\,\Gamma(1+k-n/2-i\nu)}
\biggr]
-\biggl[
c.c.
\biggr].
\label{jprod}
\end{align}
Both the $\phi$ and $|\rho|$ integrations in eq.\ (\ref{igv})
are performed term by term in the sums (\ref{jprod}).
All the $\phi$ dependence of the integrand in (\ref{igv})
is due to expression (\ref{jprod}) where in the subsequent terms
only integer powers of $\exp(i\phi)$ appear.
Thus, after the angular integration, only terms with $n+2k-2l=0$
contribute, and one of the summations in eq.\ ({\ref{jprod})
may be trivially performed giving
\begin{align}
\int_0 ^{2\pi}  d\phi\,
\biggl[ J_{n/2-i\nu}(q^{*}\rho/4)\,J_{-n/2-i\nu}(q\rho^{*}/4)
\,-\,
(-1)^{n}\,J_{-n/2+i\nu}(q^{*}\rho/4)\,J_{n/2+i\nu}(q\rho^{*}/4)
\biggr]
\nonumber \\
=
\biggl[\; 2\pi\,
%\biggl( \frac{|q||\rho|}{8}\biggr)^{-2i\nu}
\sum_{l=0}^{\infty}
\frac{ (-1)^{|n|/2}\; (|q||\rho|/8)^{4l+|n|-2i\nu}}
{\Gamma(1+l)\,\Gamma(1+l+|n|/2)\,\Gamma(1+l-i\nu)\,
 \Gamma(1+l+|n|/2-i\nu)}\biggr]
%\nonumber \\
-\biggl[c.c. \biggr].
%\hspace{5cm}
\label{bessphi}
\end{align}
Note that the odd $n$ contributions are all zero.
Using (\ref{bessphi}) and performing the remaining $|\rho|$
integration in eq.\ (\ref{igv}), we have
\begin{align}
\int_{0}^{\infty} d|\rho|\, |\rho|^{2}\, K_{0}(q_{\|}|\rho|)\,
\int_0 ^{2\pi} d\phi\,
\hspace{9cm}
\nonumber \\ \times
\biggl[
J_{n/2-i\nu}(q^{*}\rho/4)\,J_{-n/2-i\nu}(q\rho^{*}/4) \,-\,
(-1)^{n}\, J_{-n/2+i\nu}(q^{*}\rho/4)\, J_{n/2+i\nu}(q\rho^{*}/4)
\biggr]
%\hspace{6cm}
\nonumber \\
= \biggl[
\frac{4\pi}{q_{\|}^{3}}\,
%\biggl( \frac{|q|}{4q_{\|}}\biggr)^{-2i\nu}
\sum_{l=0}^{\infty} \frac{(-1)^{|n|/2}\;
\Gamma^{2}(3/2-i\nu+2l+|n|/2)\; (|q|/4q_{\|})^{4l+|n|-2i\nu}}
{\Gamma(1+l)\,\Gamma(1+l+|n|/2)\,\Gamma(1+l-i\nu)\,\Gamma(1+l+|n|/2-i\nu)}
\biggr]
%\nonumber \\
-\biggl[c.c. \biggr].
%\hspace{6cm}
\label{tauseries}
\end{align}
The obtained series is convergent for $|q|/4q_{\|}<1$, where the infinite
sum gives the value of the integral. We need to continue analytically
the result to  $|q|/4q_{\|} \geq 1$. Thus, we represent this infinite
sum of terms $\mathcal{A}(l)$ as an integral over complex $l$ using
a Sommerfeld-Watson  type transform.
A contour $\mathcal{C}_1$ in the complex $l$ plane is introduced enclosing
the half-plane $\mathcal{R}e\,(l) > -1/2$ where $\mathcal{A}(l)$ has no poles.
It is possible to construct a function $\mathcal{D}(l)$
with a pole structure and with residues such
that the contour integral of $\mathcal{D}(l)\mathcal{A}(l)/(2\pi i)$
along $\mathcal{C}_1$, evaluated using the Cauchy theorem,
reproduces the initial sum over the index $l$.
It is easy to verify that
\begin{align}
\mathcal{D}(l) \,=\,
-\pi\frac{\sin \pi i\nu}{\sin \pi l \sin \pi (l-i\nu)} \,=\,
\frac{\Gamma(l)\Gamma(1-l)\Gamma(l-i\nu)\Gamma(1-l+i\nu)}
{\Gamma(i\nu)\Gamma(1-i\nu)}
\end{align}
has the desired properties. Thus one has
\begin{align}
%\hspace{3cm} \nonumber \\
\biggl [
\frac{4\pi}{q_{\|}^{3}}
%\biggl(\frac{|q|}{4q_{\|}}\biggr)^{-2i\nu}
\sum_{l=0}^{\infty}
\frac{(-1)^{|n|/2}\;\Gamma^{2}(3/2-i\nu+2l+|n|/2)\;
(|q|/4q_{\|})^{4l+|n|-2i\nu}}
{\Gamma(1+l)\,\Gamma(1+l+|n|/2)\,\Gamma(1+l-i\nu)\,
\Gamma(1+l+|n|/2-i\nu)} \biggr]
%\nonumber \\
-\biggl[ c.c. \biggr]
%\hspace{6cm}
\nonumber \\
\,=\, \frac{4\pi}{q_{\|}^{3}}
%\biggl( \frac{|q|}{4q_{\|}}\biggr)^{-2i\nu}
\oint_{\mathcal{C}_1} \frac{dl\,\mathcal{D}(l)}{2\pi i}
\frac{(-1)^{|n|/2}\;(|q|/4q_{\|})^{4l+|n|-2i\nu}\;
\Gamma^{2}(3/2-i\nu+2l+|n|/2)
}
{\Gamma(1+l)\,\Gamma(1+l+|n|/2)\,\Gamma(1+l-i\nu)\,\Gamma(1+l+|n|/2-i\nu)}.
\label{swt1}
\end{align}
The contribution from the region of complex $l \to \infty$
vanishes in the limit, so the value of the contour
integral is given by a line integral
\begin{align}
\biggl[\frac{4\pi}{q_{\|}^{3}}
%\biggl( \frac{|q|}{4q_{\|}}\biggr)^{-2i\nu}
\sum_{l=0}^{\infty}
\frac{ (-1)^{|n|/2}\;\Gamma^{2}(3/2-i\nu+2l+|n|/2)\;
(|q|/4q_{\|})^{4l+|n|-2i\nu}}
{
\Gamma(1+l)\,\Gamma(1+l+|n|/2)\,
\Gamma(1+l-i\nu)\,\Gamma(1+l+|n|/2-i\nu)}
\biggr] - \biggl[c.c. \biggr]
\nonumber \\
\,=\, -\frac{4\pi}{q_{\|}^{3}}
%\biggl( \frac{|q|}{4q_{\|}}\biggr)^{-2i\nu}
(-1)^{|n|/2}\;\int_{-1/2-i\infty}^{-1/2+i\infty}{dl\over 2\pi i}\;
(|q|/4q_{\|})^{4l+|n|-2i\nu} \hspace{5cm}
\nonumber \\
\times
\frac{\Gamma(l-i\nu)\,\Gamma(1-l+i\nu)}{\Gamma(i\nu)\,\Gamma(1-i\nu)}\,
\frac{\Gamma(l)\,\Gamma(1-l)}{\Gamma(1+l)\,\Gamma(1+l-i\nu)}\,
\frac{\Gamma^{2}(3/2+2l+|n|/2-i\nu)}{\Gamma(1+l+|n|/2)\,\Gamma(1+l+|n|/2-i\nu)}
.
\label{swt2}
\end{align}
This form is suggestive of the $n=0$ result of \cite{BFLW}.
Substituting $s=2l+1-i\nu$, $\tau=q^{2}/4q^{2}_{\|}$ and
using the Euler $\Gamma$-function relations such as the doubling formula
$2^{2z-1}\Gamma(z)\Gamma(z+1/2)=\sqrt{\pi}\,\Gamma(2z)$
allows simplification of the integrand.
Inserting (\ref{ehat}) and (\ref{bnu}) into (\ref{igv1}),
taking into account (\ref{swt2}) and using the identity
$(-1)^{|n|/2}\,\frac{\Gamma(i\nu+|n|/2)\,\Gamma(1-i\nu-|n|/2)}
{\Gamma(i\nu)\,\Gamma(1-i\nu)}\,=\, 1$ for even $n$,
we obtain the final answer,
\begin{align}
I^{\gV}_{n,\nu}(q)\;=\;\mathcal{C}\,\as\,\frac{8\pi^2}{|q|^{3}}
\biggl( \frac{|q|^{2}}{4} \biggr) ^{i\nu}   \,
\biggl( \frac{q^*}{q}     \biggr) ^{n/2}    \,
\biggl( \frac{1}{4}       \biggr) ^{|n|/2} \,
%\frac{\Gamma(i\nu+|n|/2)\Gamma(1-i\nu-|n|/2)}
%{\Gamma(i\nu)\Gamma(1-i\nu)} \,
%\nonumber \\ \times
\frac{\Gamma(1/2-i\nu+|n|/2)}{\Gamma(1/2+i\nu+|n|/2)}
\nonumber \hspace{5cm} \\ \times
%%%%
\int_{-i\infty}^{i\infty} {ds\over 2\pi i}\,
\tau^{1/2+s+|n|/2}\,
%\nonumber \\ \times
\frac{\Gamma(1-s-i\nu)\,\Gamma(1-s+i\nu)}
{\Gamma(1-s/2-i\nu/2)\,\Gamma(1-s/2+i\nu/2)}
\nonumber \\ \times
%%%%
\frac{\Gamma^{2}(1/2+s+|n|/2)}
{\Gamma(1/2+s/2-i\nu/2+|n|/2)\,\Gamma(1/2+s/2+i\nu/2+|n|/2)}
\label{ivfinal}
\end{align}
for even $n$ and $I^{\gV} _{n,\nu}=0$ for odd $n$.
This agrees with the corresponding expression of Bartels \emph{et al.}
for $n=0$ \cite{BFLW}. The r.h.s.\ of equation (\ref{ivfinal}) is the desired
analytic continuation of the sum of the infinite power series
(c.f.\ (\ref{tauseries})) which holds for all values of $\tau$.
Introducing the notation $m=n/2$, and
using the result for $I_{n,\nu}^{qq}$, we arrive at the amplitude
\begin{align}
\mathcal{F}(z,\tau)=4\mathcal{C}\,\as^2\,\sum_{m=-\infty}^{m=\infty}
\biggl( -\frac{1}{4}\biggr)^{|m|}\int \,d\nu\,
\frac{\nu^{2}+m^{2}}{(\nu^{2}+(m-1/2)^{2})(\nu^{2}+(m+1/2)^{2})}
\; e^{\chi_m(\nu)z} \
\nonumber \\
\times
\int_{-i\infty}^{i\infty}\frac{ds}{2\pi i}\, \tau^{1/2+s+|m|}\,
\frac{\Gamma(1-s-i\nu)\,\Gamma(1-s+i\nu)}
     {\Gamma(1-s/2-i\nu /2)\,\Gamma(1-s/2+i\nu /2)}
\nonumber \\
\times
\frac{\Gamma^{2}(1/2+s+|m|)}
{\Gamma(1/2+s/2-i\nu/2+|m|)\,\Gamma(1/2+s/2+i\nu /2+|m|)},
\label{master}
\end{align}
which is equal to
\begin{align}
\mathcal{F}(z,\tau)=\frac{8\mathcal{C}\,\as^2}{\pi}
\sum_{m=-\infty}^{m=\infty} (-1)^{m}
\int \,d\nu\,
\frac{\nu^{2}+m^{2}}{(\nu^{2}+(m-1/2)^{2})(\nu^{2}+(m+1/2)^{2})}
\; e^{\chi_m(\nu)z} \
\nonumber \\
\times
\int_{-i\infty}^{i\infty}\frac{ds}{2\pi i}\,
\biggl( {\tau \over 4} \biggr) ^{1/2+s+|m|}\,
\frac{\Gamma(1/2-s/2+i\nu/2)\,\Gamma^{2}(1/2+s+|m|)\,\Gamma(1/2-s/2-i\nu/2)}
{\Gamma(1/2+s/2-i\nu/2+|m|)\,\Gamma(1/2+s/2+i\nu /2+|m|)}.
\label{master2}
\end{align}
Note that the $m$ contribution is equal to the $-m$ contribution.

%%%%%%%%%%%%%%%%%%%%%%%%%%%%%%%%%%%%%%%%%%%%%%%%%%%%%%%%%%%%%%%%%%%%%%%%%%%
\section{Properties of the solutions}\label{sec5}

The form of the solution to the BFKL equation given by eq.\ (\ref{master2})
is simple enough to perform extended studies of the amplitude, including
the impact of higher conformal spins. The remaining complex integrations
over $s$ and $\nu$ are performed numerically. In the analysis we
take the real photon case, $Q_{\gamma}^2=0$.
First, we set $z=0$ in which case the LO BFKL amplitude is
described by a simple two-gluon exchange,
\begin{equation}
\mathcal{F}(z=0,\tau) = \mathcal{C}\;\as^2
\biggl( {4\tau ^2 \over 1-\tau^2 } \biggr)\;
\ln\, \biggl( { (1+\tau)^2 \over 4\tau } \biggr).
\label{2g}
\end{equation}

\begin{figure}
\begin{center}
\epsfig{width=0.65 \columnwidth, file=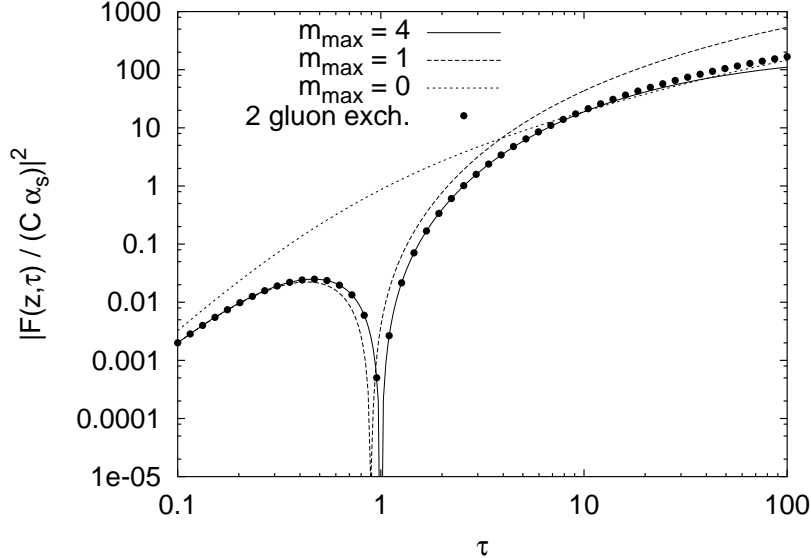} \\
\end{center}
\caption{\em
The amplitude squared for the diffractive heavy vector meson
production off a quark at zero rapidity ($z=0$):
the exact two-gluon exchange result (dots) compared
with sums of contributions up to conformal spin
$n_{\mathrm{max}} = 2m_{\mathrm{max}}$ (lines)}
\label{fig2}
\end{figure}
A valuable cross check of our calculation is to investigate how the
two-gluon exchange amplitude builds up when subsequent higher conformal
spin components are being added. In fig.\ \ref{fig2} we show curves 
corresponding to
$|\mathcal{F}(z,\tau)/(\mathcal{C}\,\as)|^2$ approximated by partial sums
over $m$ in (\ref{master2}) up to $m=m_{\mathrm{max}} =0,1,4$ and compare
them with the amplitude given by eq.\ (\ref{2g}). It is clear that results
coming from the two approaches agree. The sum over conformal spins
converges quickly to the exact result, however more terms are needed for
increasing $\tau$. Note, that the $m=n=0$ component and the exact result
have about the same absolute values for $\tau \gg 1$ but the signs of the
amplitudes are opposite.

\begin{figure}
\begin{center}
{\large a)}\hspace{5mm} \epsfig{width=0.65 \columnwidth, file=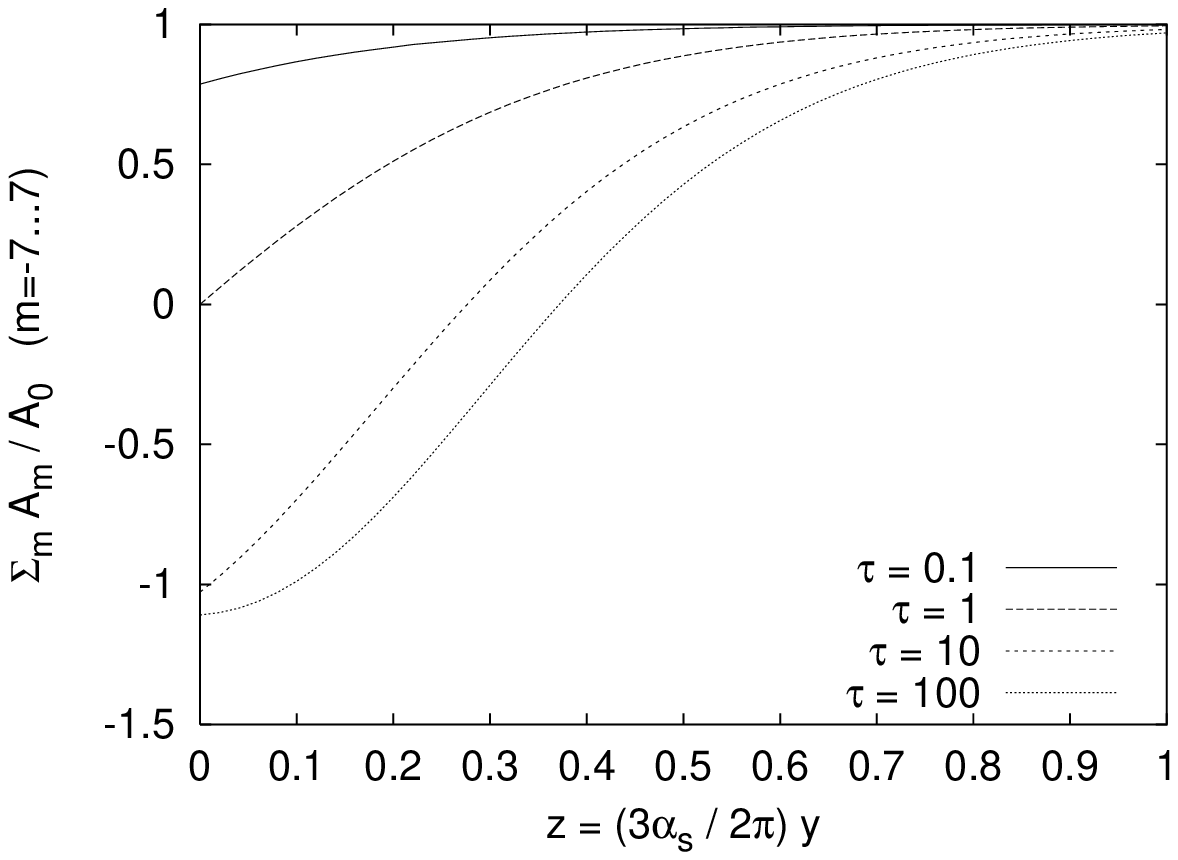}
\hspace{10mm} \\[14mm]
{\large b)}\hspace{5mm} \epsfig{width=0.65 \columnwidth, file=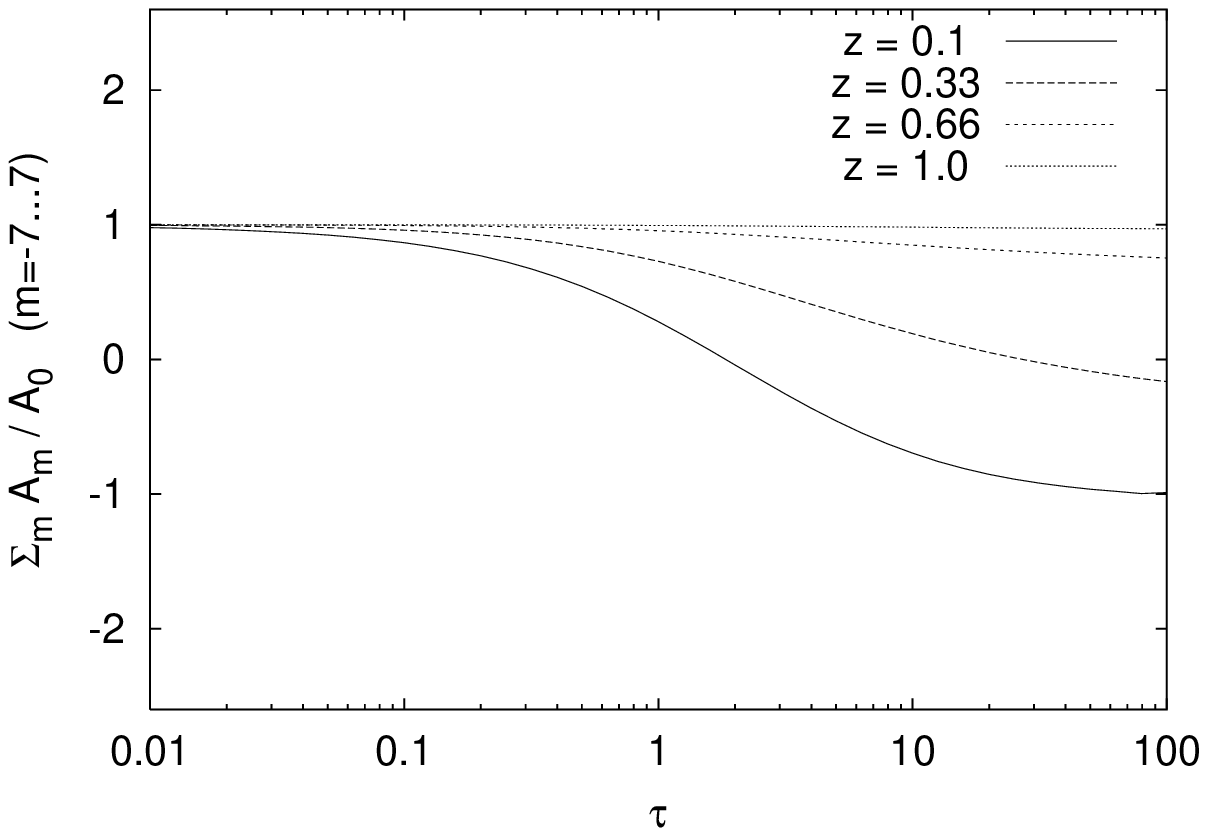}
\hspace{10mm} \\
\end{center}
\caption{\em Comparison of the BFKL amplitude approximated by a partial
sum in eq.\ (\ref{master2}) up to conformal spin $|n|=|2m|=14$ with
the leading conformal spin result, $n=0$. Dependencies of the ratio
are given as a function of a) $z$ and b) $\tau$ respectively.
}
\label{fig3}
\end{figure}

\begin{figure}
\begin{center}
\epsfig{width=0.65 \columnwidth, file=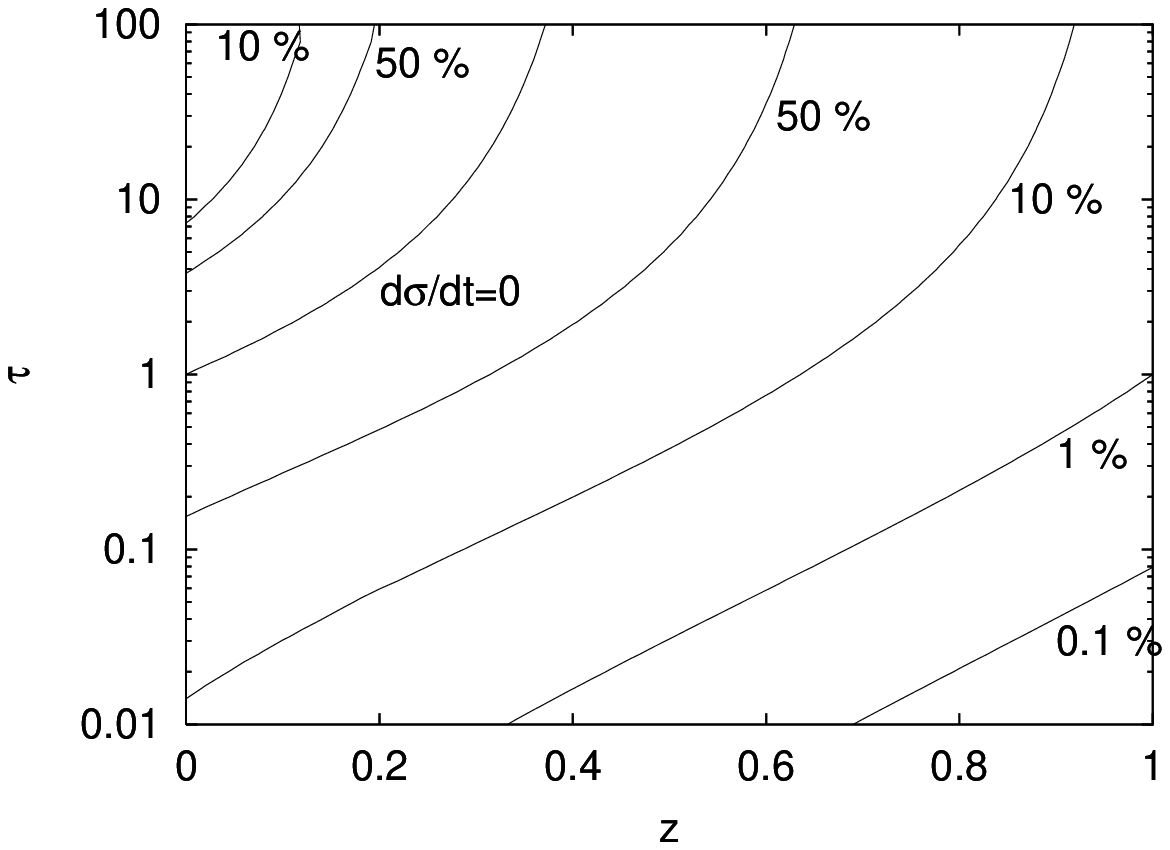} \\
\end{center}
\caption{\em Contour plot showing the relative error $\mathcal{E}$
of the  cross-section obtained in the $n=0$ approximation to formula
(\ref{master2}) in the $(z,\tau)$ plane. In this plot
$\mathcal{E}<0$. The contour labelled $d\sigma/dt = 0$
corresponds to values of $z$ and $\tau$ for which the exact
cross-section vanishes.}
\label{contour}
\end{figure}

Having checked this, we are in a position to study the importance of higher
conformal spins at $z > 0$.
Thus, in fig.\ \ref{fig3} the ratio of $\mathcal{F}(z,\tau)$
(with $m_{\mathrm{max}}=7$) to the $m=0$ component is plotted
for various $z$ as a function of $\tau$ and
for various $\tau$ as a function of $z$.
The relative importance of the higher conformal spins for
the cross-section may be read out from fig.~\ref{contour}.
The contours show constant values of
\begin{equation}
\mathcal{E} =
{ d\sigma/dt|_{\mathrm{exact}} - d\sigma/dt|_{n=0}
\over d\sigma/dt|_{n=0} },
\label{accur}
\end{equation}
giving therefore the relative error of the conventional
leading conformal spin approximation in the $(z,\tau)$ plane.
As expected, the correction due to $m \neq 0 $ components
decreases with increasing $z$ (or $y$) and increases with
increasing $\tau$ (or $|t|$). Note also the line
$\,d\sigma / dt = 0\,$ on which the complete amplitude
changes sign, leading to a dip in the cross-section
$\,d\sigma/dt|_{\mathrm{exact}}$.
Such a dip does not appear in the leading
conformal spin approximation.

The analytical results may also be used to test the method and
approximations used in the numerical approach to the non-forward
BFKL equation \cite{PSIPSI,EIM}. In fig.\ \ref{fig5}, a comparison
between results obtained in these two frameworks is given.
With good accuracy the numerical solution coincides with
the analytical one.
%%% comm 4
At larger values of $z$ and $\tau$ a discrepancy between
the two solutions appears. This is caused by an upper cut-off
$K_F = 10^5$~GeV imposed on the gluon transverse momenta, 
which is necessary for the numerical approach. The very large gluon momenta 
that are affected by this procedure contribute significantly 
to the amplitudes when the momentum transfer and/or the rapidity 
is very large.
%%%%

\begin{figure}
%%%%%%%% fig 5
\begin{center}
a) \hspace{5mm} \epsfig{width=0.65 \columnwidth, file=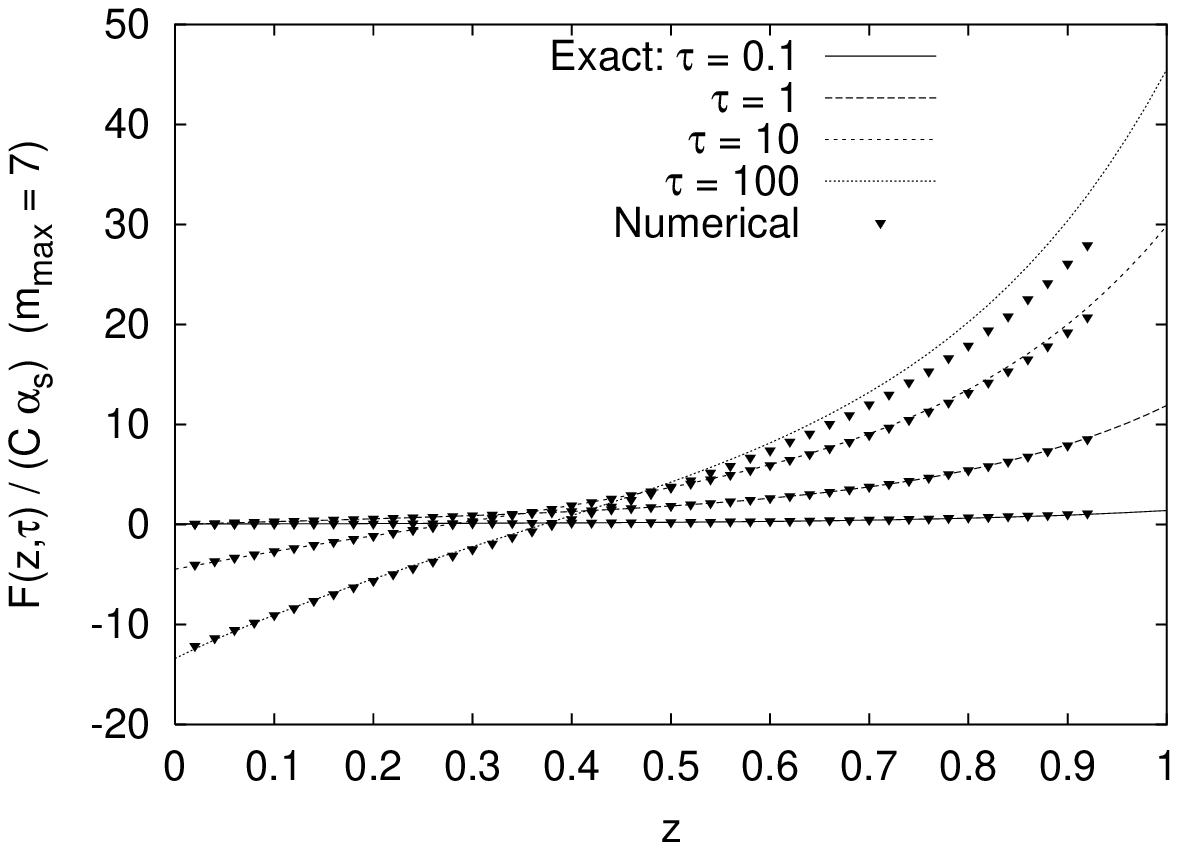} \hspace{1cm}\\
b) \hspace{5mm} \epsfig{width=0.65 \columnwidth, file=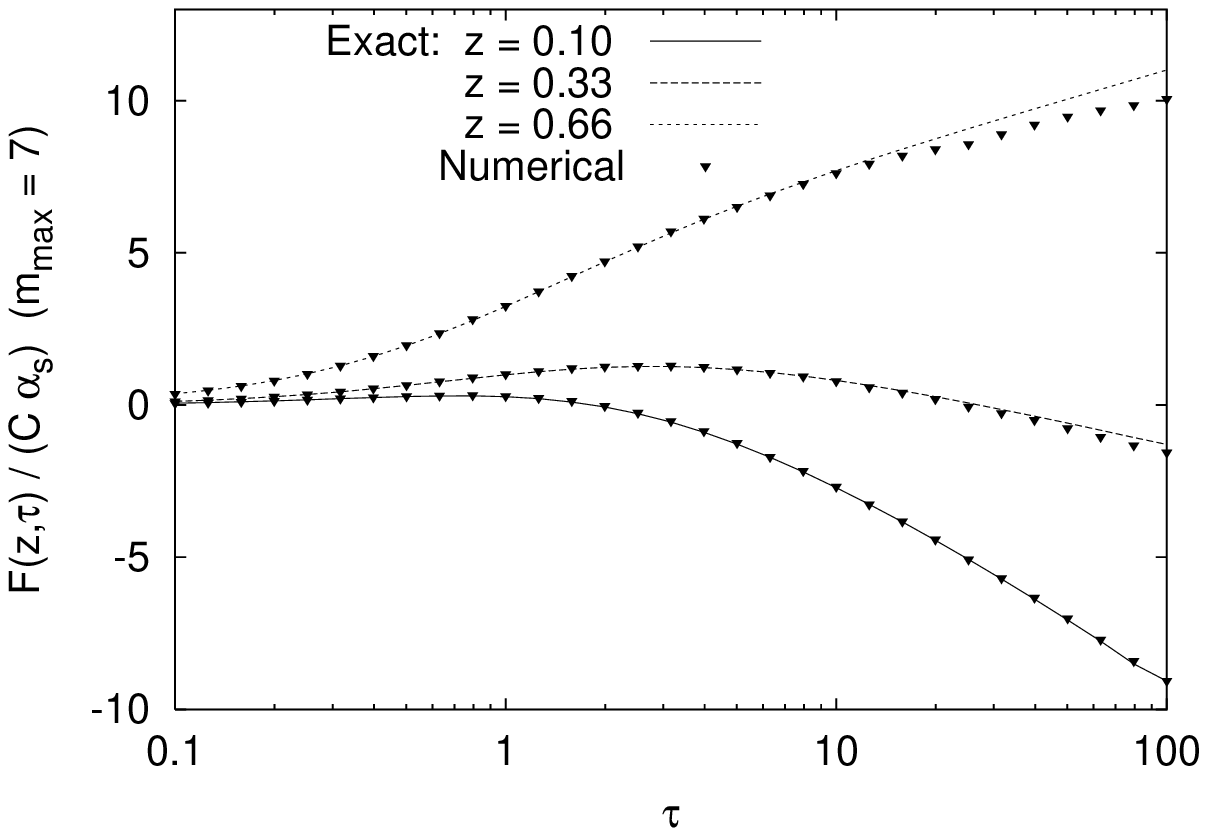} \hspace{1cm}\\
\end{center}
\caption{\em Comparison of analytical and numerical results shown as a function
of a) $z$ for various $\tau$ b) $\tau$ for various $z$. }
\label{fig5}
\end{figure}

The non-leading corrections \cite{BFKLNL} to the BFKL equation have
a large impact on the rapidity dependence of the cross-sections, as
already discussed. At high rapidities, this leads to dramatic effects
in the magnitude of the cross-section \cite{PSIPSI}. Therefore
it is important to incorporate those non-leading effects in
the analysis. In this case, the conformal symmetry of the kernel
is broken and no exact analytical approach is known yet.
Fortunately, using the more straightforward numerical method
one may obtain a solution to the BFKL equation beyond the leading
logarithmic approximation (BFKL~LL+NL), given by eq.\ (\ref{bfklkc}).
\begin{figure}
\begin{center}
a) \hspace{5mm} \epsfig{width=0.65 \columnwidth, file=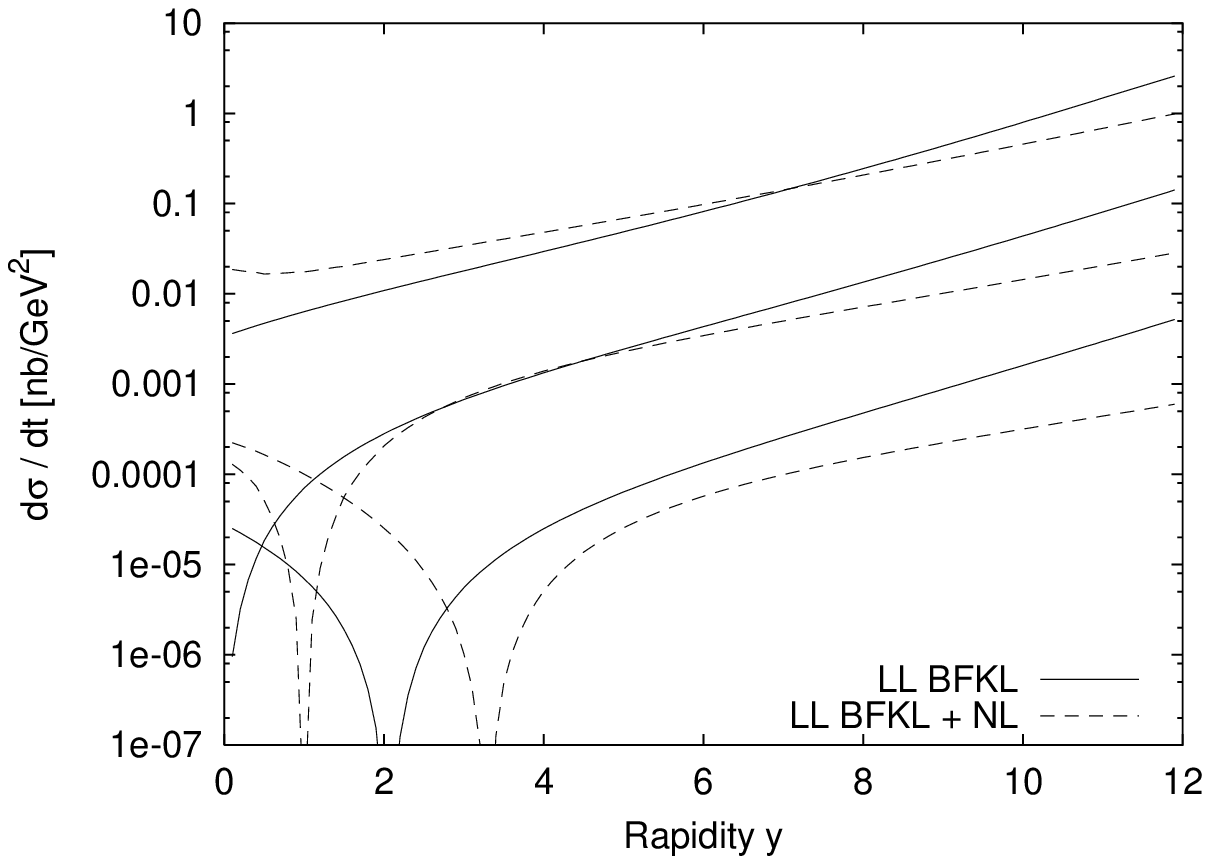} \hspace{1cm}\\
b) \hspace{5mm} \epsfig{width=0.65 \columnwidth, file=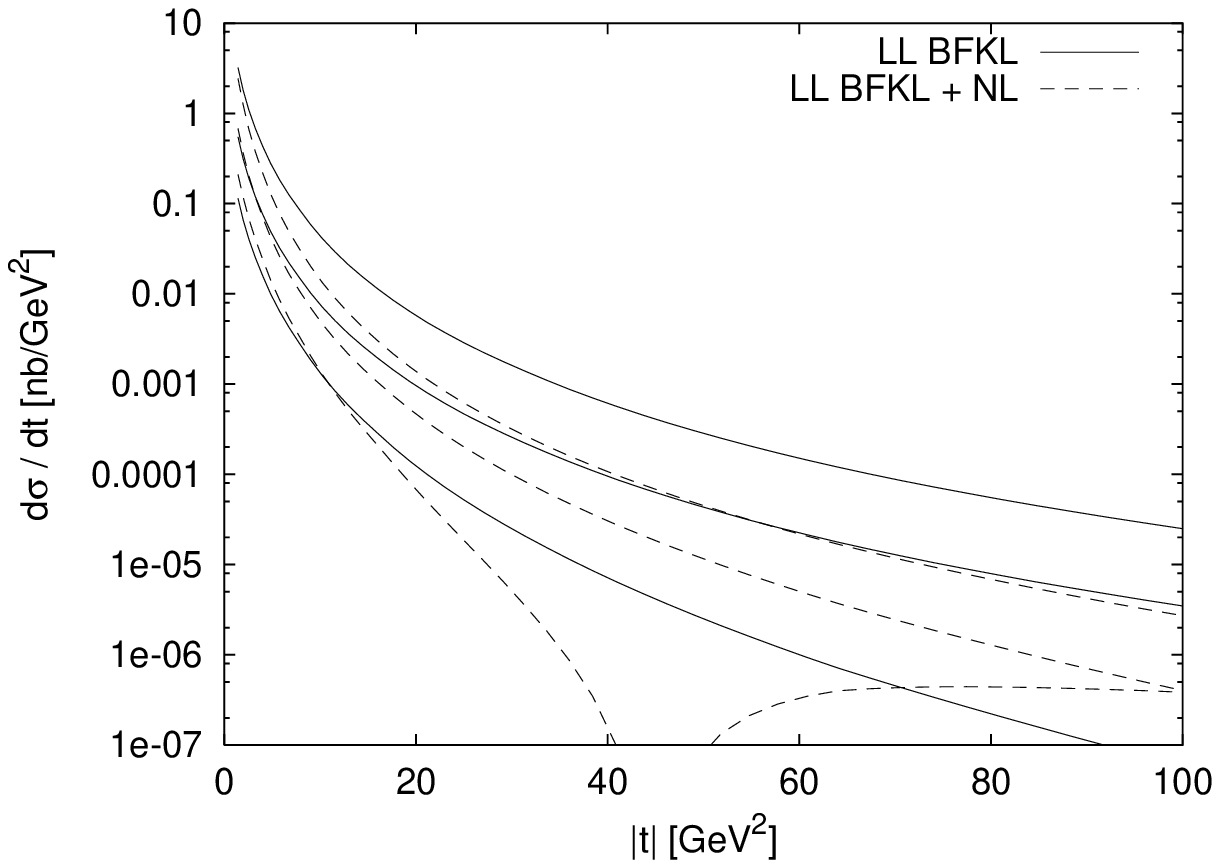} \hspace{1cm}\\
\end{center}
\caption{\em The cross-sections $d\sigma/dt (y,|t|)$
for diffractive $J/\psi$ production off a quark shown as a function of
a) rapidity $y$ for $|t|=3$~GeV$^2$ (uppermost curves),
10~GeV$^2$ (in the middle) and 30~GeV$^2$ (lowermost curves),
and b) momentum transfer $|t|$ for $y=4$ (lowermost curves),
$y=7$ (in the middle) and $y=10$ (uppermost curves).
Solid curves are obtained from the LL BFKL equation and
dashed ones from the BFKL LL+NL equation
}
\label{fig6}
\end{figure}
In fig.\ \ref{fig6} the rapidity $y=\ln (1/x)$ (c.f.\ (\ref{bfkl}), (\ref{bfklkc}))
and $|t|$ dependencies of the cross-section $d\sigma/dt\,(y,|t|)$
for $J/\psi$ production off a quark are shown. The cross-sections
obtained from the LL~BFKL and BFKL~LL+NL are compared.
At the leading logarithmic accuracy the value of the fixed coupling
$\alpha_s$ is not constrained and we choose $\as = 0.17$.
With this choice, the LL and LL+NL results in the studied window
of $y$ and $|t|$  have a similar overall normalization.
In the BFKL LL+NL equation (\ref{bfklkc}), we take $s_0=0.5$~GeV$^2$,
and the scales of the running coupling according to (\ref{scales}).
It may be seen in fig.\ \ref{fig6}a, that the increase for large $y$
is less steep (i.e.\ the intercept is smaller) for
the LL+NL case, in spite of taking a rather small value of
$\as$ in the LL~BFKL equation. To be precise, the pomeron
intercept is $\alpha_P = 1.53$ for the LL~BFKL curves
and $\alpha_P \simeq 1.3$ for the non-leading solution.
It is clear, that the non-leading prediction is much
closer to the experimental estimates of  $\alpha_P \simeq 1.2$
in the vector meson production process.

%%%%%%% fig. 7
\begin{figure}[!th]
\begin{center}
\epsfig{width=0.65 \columnwidth, file= 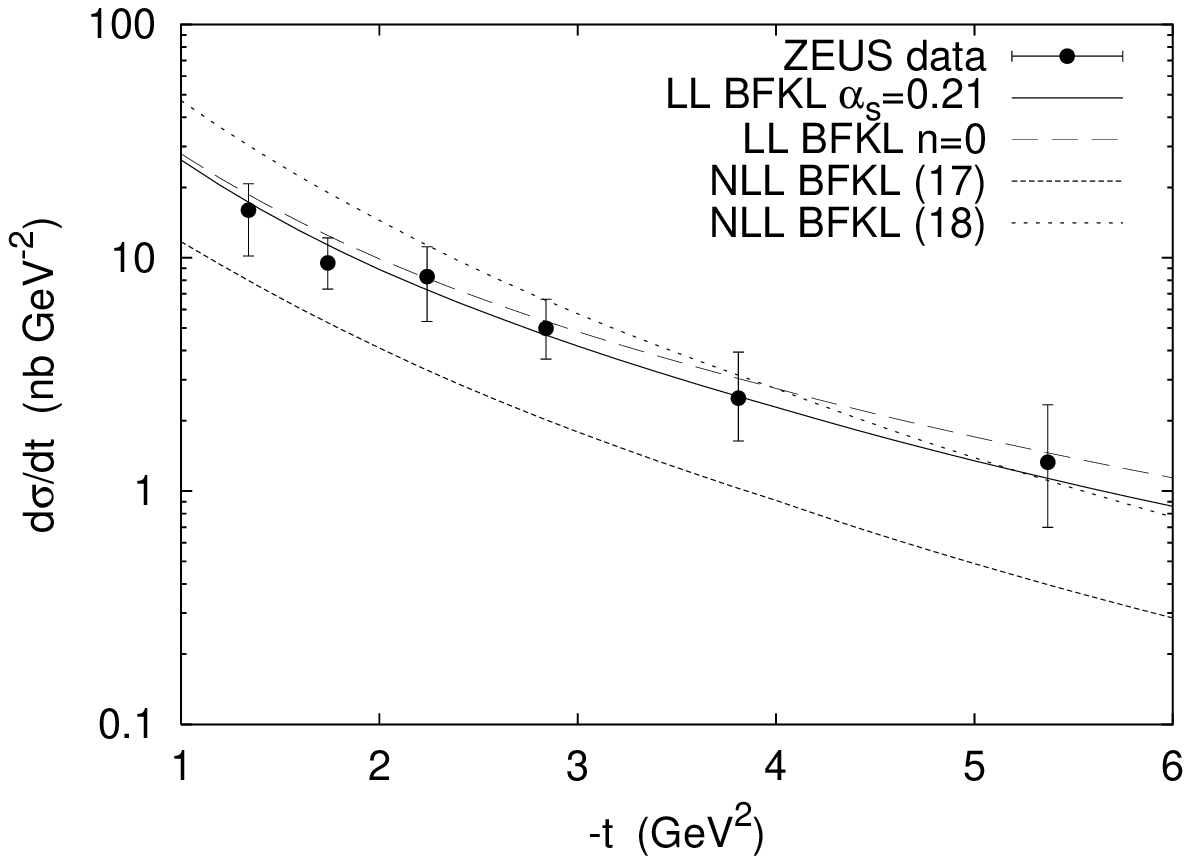} \\
\end{center}
\caption{\em The cross-section $d\sigma/dt$
for diffractive $J/\psi$ photo-production off proton
shown as a function of the momentum transfer $|t|$.
ZEUS data are compared with the theoretical results from:
LL~BFKL (continuous line), the leading conformal spin
approximation to LL~BFKL amplitude (dashed line)
and the BFKL equation with non-leading corrections
(dotted lines where lower and upper lines correspond
to the choices of scales given by (\ref{scales})
and (\ref{scales2}) respectively). A correlated
10\% uncertainty of the normalization of data
points is not included into the error bars.
}
\label{fig7}
\end{figure}

Thus, when both the normalization (which also depends  on $\as$) and
the rapidity dependence are taken into account, the need for NL
effects in the  BFKL kernel should become visible.
Comparison of the shapes in $|t|$, given in fig.\ \ref{fig6}b, demonstrates that
they are similar in both cases, especially in the low~$|t|$ range.
The LL+NL curves are steeper because of the running of $\alpha_s$.
To conclude, the main impact of the non-leading corrections
seems to be a reduction of the pomeron intercept.

%%%%%%%%%%%%%%%%%%%%%%%%%%%%%%%%%%%%%%%%%%%%%%%%%%%%%%%%%%%%%%%%%%%%%%%%%%%
\section{Comparison with data}\label{sec6}

The results of the model calculation described in the previous
sections may be compared to the ZEUS data \cite{ZEUS}
on diffractive $J/\psi$ photoproduction.
In this measurement, the photon virtuality $Q_{\gamma}^2
\simeq 0$ and the photon-proton collision energy
is in the range $80~\mbox{GeV} < W < 120~\mbox{GeV}$.
In fig.\ \ref{fig7} the data are shown together with the theoretical curves
obtained from eq.\ (\ref{dsdtgp}) with various models.
%%% comm 5
These curves were obtained using the CTEQ5L parton distributions
\cite{CTEQ}.
%%%
The continuous curve
is given by the analytical solution of the BFKL equation (\ref{master2})
with $\alpha_s = 0.21$ and all conformal spins included, and the dashed one
corresponds to the leading conformal spin ($n=2m=0$) in (\ref{master2}).
The dotted curves are obtained from the BFKL equation with non-leading
effects (\ref{bfklkc}). For the lower curve, the strong coupling constant
in the impact factors is evaluated at the scales given by  (\ref{scales}),
and for the upper curve we use the values given by (\ref{scales2})
as described in Sec.\ \ref{sec3}.

The LL~BFKL results fit the data very well and the difference between the
leading conformal spin and the full solution is small, although the
discrepancy increases with~$t$. The non-leading BFKL results are in
rather good agreement with the data when low scales of $\alpha_s$ are
chosen (\ref{scales2}), but underestimate the data when the most
natural choice (\ref{scales}) of scales is made.
Thus, in formulating predictions extrapolating beyond
the currently measured kinematical window, we will
use the data-guided option (\ref{scales2}).
%
% comm 3b
We have also checked, that the sensitivity of the normalization
of the non-leading BFKL results to the value of $s_0$ is
relatively large within the considered window:\ for the scale choice 
(\ref{scales}), it increases by a factor of two when $s_0$ 
decreases from $0.5$~GeV$^2$ to $0.1$~GeV$^2$.
The $t$-dependence is only slightly steeper at low $t$ for 
the latter case. Therefore, the shape of the cross-section is stable against
variation of details, while the normalization is more uncertain, and
critical tests of BFKL should be based on the shape rather than the 
normalization of the cross-section.

Recall that an important feature of the non-leading BFKL solution is the
emerging value of the pomeron intercept of about~1.3, to be compared with
the LL~BFKL value $\alpha_P = 1.56$ for $\alpha_s = 0.21$. Thus, we expect
that the LL~BFKL should overestimate significantly the cross-sections for
larger average collision energies~$W$. An interesting experimental
verification of the impact of non-leading corrections on the scattering
amplitudes could be provided by performing analogous measurements at higher
energies~$W$. In fig.\ \ref{fig8} theoretical estimates from
LL~BFKL and non-leading BFKL are shown for the photon-proton
collision energy $W=100 $ and 200~GeV, for a wide $t$~range.
The parameters are adjusted to give the best fits
of the presently available ZEUS data, that is $\as = 0.21$ for LL~BFKL,
$s_0 = 0.5$~GeV$^2$ and the running coupling being taken at
scales~(\ref{scales2}). 
One may see that although the inclusion of non-leading corrections lowers
the expected value of the cross-sections at $W=200$~GeV
it may be insufficient to discriminate between the models.
We have checked that the impact of higher conformal spins may
be safely neglected at $W=200$~GeV and $|t| < 10$~GeV$^2$.

\begin{figure}
\begin{center}
a) \hspace{5mm} \epsfig{width=0.65 \columnwidth, file=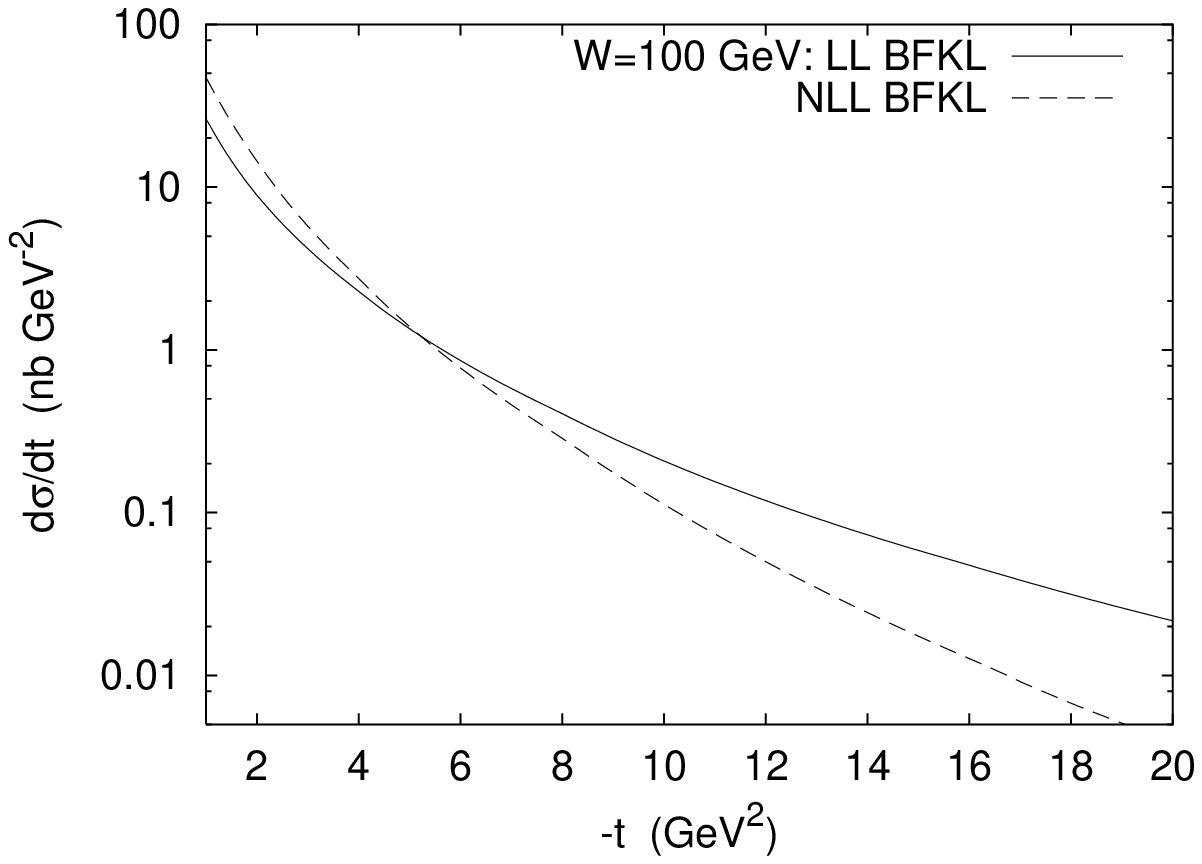} \hspace{1cm}\\
b) \hspace{5mm} \epsfig{width=0.65 \columnwidth, file=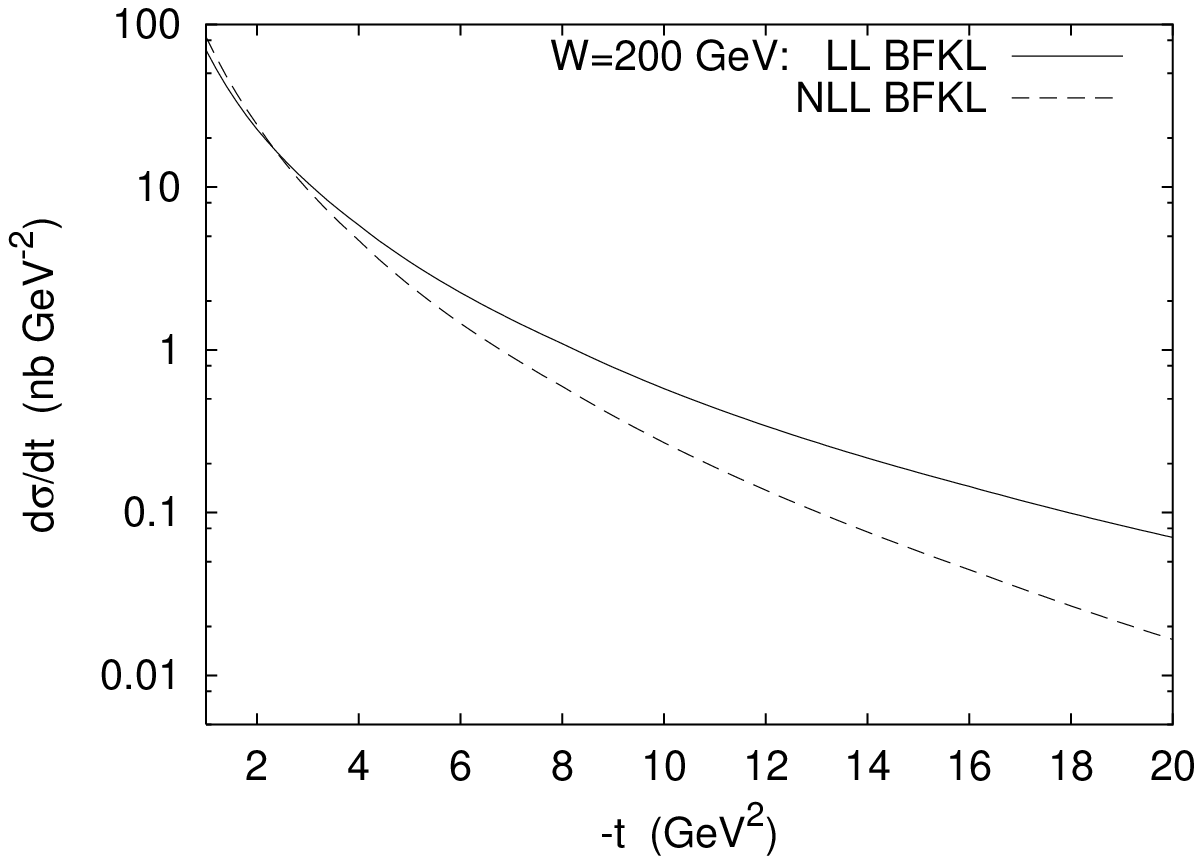} \hspace{1cm}\\
\end{center}
\caption{\em The cross-sections $d\sigma/dt$
for diffractive $J/\psi$ photo-production off proton
shown as a function of the momentum transfer $|t|$
for $\gamma p$ collision energy a)~$W=100$~GeV and
b)~$W=200$~GeV. The continuous and dotted lines represent
the LL~BFKL and BFKL with non-leading corrections
results respectively.}
\label{fig8}
\end{figure}

Note that the cross-sections fall off much steeper
with increasing~$t$ when the non-leading corrections are included
(see fig.\ \ref{fig8}~a,b). This is mostly because of the
running of the coupling with the energy scale
related to the momentum transfer.

%%%%%%%%%%%%%%%%%%%%%%%%%%%%%%%%%%%%%%%%%%%%%%%%%%%%%%%%%%%%%%%%%%%%%%%%%%%
\section{Conclusions}\label{sec7}
In this paper an analysis was performed of BFKL amplitudes for
diffractive heavy vector meson photoproduction at large momentum transfer.
We obtained an explicit complete solution to the leading-logarithmic BFKL
equation describing this process. The novel feature of our approach is the
inclusion of terms subleading at very high rapidity, corresponding to
higher conformal spins in Lipatov's expansion of the BFKL amplitude.
These subleading effects were found to reduce theoretical expectations for
the cross-sections by about $10\%$ in the kinematical window currently
probed by experiments on the $J/\psi$ production.
This result gives a firmer ground for the previous results,
based on the leading conformal spin approximation.
The relative importance of higher conformal spins increases,
however, with decreasing collision energy or increasing
momentum transfer, as shown in fig.~\ref{contour}.

Also non-leading corrections to BFKL equations were taken into account
phenomenologically by using the running coupling constant and applying the
so-called consistency constraint in the BFKL kernel. In this case a
numerical method was used to solve the equation. The main influence of
non-leading corrections was found to be a reduction of the hard pomeron
intercept to about~1.3, close to the value determined from experiment.

Results obtained from both approaches were compared to the experimental
data on the $t$-dependent differential cross-section for $J/\psi$
photoproduction at $\gamma p$ collision energy $W \sim 100$~GeV. In both
cases a good fit was obtained, although we found that the cross-section
grows much slower with rapidity when non-leading corrections are included,
leading to a discrepancy from the LL~BFKL results increasing with rapidity.
The ratio of the differential cross-sections at $W=200$~GeV and $W=100$~GeV
may be used to find the influence of non-leading corrections to
the BFKL equation if the data are accurate enough.

To summarize, we provide more insight into the BFKL mechanism of diffractive
heavy vector meson production and confirm that the available
data are consistent with BFKL expectations.

%%%%%%%%%%%%%%%%%%%%%%%%%%%%%%%%%%%%%%%%%%%%%%%%%%%%%%%%%%%%%%%%%%%%%%%%%%%
\section*{Acknowledgements}
We are grateful to Katarzyna Klimek, Malcolm Derrick and the
ZEUS collaboration, for their interest and for providing us with
their data. We thank Jeff Forshaw and Gunnar Ingelman for the support.
This study was supported in part by the Swedish
Research Council, by the Polish Committee for Scientific
Research (KBN) grant no. 5P03B~14420, and
by PPARC studentship PPA/S/S/2000/03130.

%%%%%%%%%%%%%%%%%%%%%%%%%%%%%%%%%%%%%%%%%%%%%%%%%%%%%%%%%%%%%%%%%%%%%%%%%%%

%%%%%%%%%%%%%%%%%%%%%%%%%%%%%%%%%%%%%%%%%%%%%%%%%%%%%%%%%%%%%%%%%%%%%%%%%%%


\begin{thebibliography}{99}

\bibitem{BFKL}
L.~N.~Lipatov,
Sov.\ J.\ Nucl.\ Phys.\  {\bf 23} (1976) 338;
%%CITATION = SJNCA,23,338;%%
%
E.~A.~Kuraev, L.~N.~Lipatov and V.~S.~Fadin,
Sov.\ Phys.\ JETP {\bf 44}, 443 (1976); {\it ibid.\ } {\bf 45} (1977) 199;
%%CITATION = SPHJA,44,443;%%
%%CITATION = SPHJA,45,199;%%
%
I.~I.~Balitsky and L.~N.~Lipatov,
Sov.\ J.\ Nucl.\ Phys.\  {\bf 28} (1978) 822.
%%CITATION = SJNCA,28,822;%%


\bibitem{Lipatov}
L.~N.~Lipatov,
Sov.\ Phys.\ JETP {\bf 63} (1986) 904;
Phys.\ Rep.\  {\bf 286} (1997) 131.
%%CITATION = HEP-PH 9610276;%%


\bibitem{FR}
J.~R.~Forshaw and M.~G.~Ryskin,
Z.\ Phys.\ {\bf C68} (1995) 137.
%%CITATION = HEP-PH 9501376;%%

\bibitem{BFLW}
J.~Bartels, J.~R.~Forshaw, H.~Lotter and M.~W\"usthoff,
Phys.\ Lett.\ {\bf B375} (1996) 301.
%%CITATION = HEP-PH 9601201;%%


\bibitem{ZEUS}
S.~Chekanov {\it et al.}  [ZEUS Collaboration],
hep-ex/0205081, Eur.~Phys.~J.~C, in press.
%%CITATION = HEP-EX 0205081;%%



\bibitem{FP}
J.~R.~Forshaw and G.~Poludniowski,
hep-ph/0107068, Eur.~Phys.~J.~C, in press.
%%CITATION = HEP-PH 0107068;%%



\bibitem{MT}
A.~H.~Mueller and W.-K.~Tang,
Phys.\ Lett.\ {\bf B284} (1992) 123.
%%CITATION = PHLTA,B284,123;%%


\bibitem{MMR}
 L.~Motyka, A.~D.~Martin and M.~G.~Ryskin,
Phys.\ Lett.\ B {\bf 524} (2002) 107.
%%CITATION = HEP-PH 0110273;%%


\bibitem{BFKLNL}
V.~S.~Fadin and L.~N.~Lipatov,
Phys.\ Lett.\ {\bf B429} (1998) 127;
%%CITATION = HEP-PH 9802290;%%
M.~Ciafaloni and G.~Camici,
Phys.\ Lett.\ {\bf B430} (1998) 349.
%%CITATION = HEP-PH 9803389;%%


\bibitem{EIM}
R.~Enberg, G.~Ingelman and L.~Motyka,
Phys.\ Lett.\ {\bf B524} (2002) 273.
%%CITATION = HEP-PH 0111090;%%


\bibitem{LVM}
R.~Enberg, J.~Forshaw, L.~Motyka and G.~Poludniowski, in preparation.
%%CITATION = NONE;%%


\bibitem{BFLLR}
J.~Bartels, J.~R.~Forshaw, H.~Lotter, L.~N.~Lipatov, M.~G.~Ryskin and
M.~W\"usthoff,
Phys.\ Lett.\ {\bf B348} (1995) 589.
%%CITATION = HEP-PH 9501204;%%

\bibitem{GLUs0}
J.~M.~Cornwall,
Phys.\ Rev.\  {\bf D26} (1982) 1453;
%%CITATION = PHRVA,D26,1453;%%
C.~Alexandrou, P.~de Forcrand and E.~Follana,
Phys.\ Rev.\  {\bf D63} (2001) 094504.
%%CITATION = HEP-LAT 0008012;%%


\bibitem{Latt1}
C.~Alexandrou, P.~de Forcrand and E.~Follana,
%``The gluon propagator without lattice Gribov copies,''
Phys.\ Rev.\ {\bf D63} (2001) 094504;
%[arXiv:hep-lat/0008012].
%%CITATION = HEP-LAT 0008012;%%
%\bibitem{Latt2}
D.~B.~Leinweber, J.~I.~Skullerud, A.~G.~Williams and C.~Parrinello  [UKQCD
                  collaboration],
%``Gluon propagator in the infrared region,''
Phys.\ Rev.\ {\bf D58} (1998) 031501.
%[arXiv:hep-lat/9803015].
%%CITATION = HEP-LAT 9803015;%%

\bibitem{DucatiSauter}
M.~B.~Gay Ducati and W.~K.~Sauter,
%``Non-perturbative gluons in diffractive photo-production of J/psi,''
Phys.\ Lett.\ B {\bf 521} (2001) 259.
%[arXiv:hep-ph/0110162].
%%CITATION = HEP-PH 0110162;%%

\bibitem{Ryskin}
M.~G.~Ryskin,
Z.\ Phys.\ C {\bf 57} (1993) 89.
%%CITATION = ZEPYA,C57,89;%%


\bibitem{FKS}
L.~Frankfurt, W.~Koepf and M.~Strikman,
%``Diffractive heavy quarkonium photo- and electroproduction in QCD,''
Phys.\ Rev.\ {\bf D57} (1998) 512.
%[arXiv:hep-ph/9702216].
%%CITATION = HEP-PH 9702216
%

\bibitem{IKSS}
D.~Y.~Ivanov, R.~Kirschner, A.~Sch\"afer and L.~Szymanowski,
%``The light vector meson photoproduction at large t,''
Phys.\ Lett.\ {\bf B478} (2000) 101
[Erratum-ibid.\ {\bf B498} (2001) 295].
%[arXiv:hep-ph/0001255].
%%CITATION = HEP-PH 0001255;%%


\bibitem{KC}
B.~Andersson, G.~Gustafson, H.~Kharraziha and J.~Samuelsson,
Z.\ Phys.\ {\bf C71} (1996) 613;
%%CITATION = ZEPYA,C71,613;%%
J.~Kwieci\'{n}ski, A.~D.~Martin and P.~J.~Sutton,
Z.\ Phys.\ {\bf C71} (1996) 585.
%%CITATION = HEP-PH 9602320;%%


\bibitem{COLLINEAR}
G.~P.~Salam,
JHEP {\bf 9807} (1998) 019;
%%CITATION = HEP-PH 9806482;%%
M.~Ciafaloni, D.~Colferai and G.~P.~Salam,
Phys.\ Rev.\  {\bf D60} (1999) 114036;
%%CITATION = HEP-PH 9905566;%%
G.~P.~Salam,
Acta Phys.\ Polon.\ {\bf B30} (1999) 3679.
%%CITATION = HEP-PH 9910492;%%

\bibitem{KMSTAS}
J.~Kwieci\'{n}ski, A.~D.~Martin and A.~M.~Sta\'{s}to,
Phys.\ Rev.\ {\bf D56} (1997) 3991.
%%CITATION = HEP-PH 9703445;%%

\bibitem{PSIPSI}
J.~Kwieci\'{n}ski and L.~Motyka,
Phys.\ Lett.\ {\bf B438} (1998) 203.
%%CITATION = HEP-PH 9806260;%%


\bibitem{KMDT}
J.~Kwieci\'{n}ski and L.~Motyka,
Phys.\ Lett.\ {\bf B462} (1999) 203.
%%CITATION = HEP-PH 9905567;%%


\bibitem{NP}
H.~Navelet and R.~Peschanski,
Nucl.\ Phys.\ B {\bf 507} (1997) 353.
%%CITATION = HEP-PH 9703238;%%


\bibitem{CTEQ}
H.\ L.\ Lai {\it et al.\/}, Eur.\ Phys.\ J.\ C {\bf 12} (2000) 375
%%CITATION = HEP-PH 9903282;%%



\end{thebibliography}
\end{document}